# Thermodynamic Theory of Sintering and Swelling

Yuri Kornyushin

Maître Jean Brunschvig Research Unit,
Chalet Shalva,
Randogne, CH-3975

The General Thermodynamic Theory of Sintering, formulated by the author in 1998 is given. This theory is applied to the problem of swelling of materials under conditions of radiation. Driving forces, caused by the presence of the evolution of heat in the volume of a sample (electric contact, hf, inductive heating or penetrating radiation, e.g., neutrons could be the sources of the heat in the bulk of a sample) are considered. The influence of these driving forces on sintering, structure and properties is discussed. The role of mobile and immobile dislocations, grain boundaries, and pores is regarded. Cycling and pulsing regimes of sintering are investigated. A mesoscopic approach, described in present paper, which had been used for years to solve sintering problems, also could help solving problems of nucleation, decomposition, and other problems in the theory of alloys. Described relaxation time technique could help simplifying calculations and give more clear physical picture. Bulks heating driving forces, important for small size objects of microelectronics or/and heat evolving elements in nuclear reactors, are important also for the problems of nucleation and decomposition in these objects.

## 1. INTRODUCTION

Directions and rates of diffusional processes of mass transfer are important constituents of the process of sintering. The thermodynamic driving forces and kinetic characteristics of a system determine them. Kinetic characteristics include magnitudes of kinetic coefficients, e.g., diffusion coefficient, and conditions of realization of different mass transfer mechanisms. A vacancy mechanism of mass transfer is a regular one in crystalline materials. The mean free path of the excess vacancy, which is determined by the defect structure of a material and by the distance from a free surface, governs essentially the rates of diffusional processes.

The process of sintering is usually considered to consist of three parts [1]. The initial stage of sintering is usually regarded to be determined by the processes of connection of separate particles, and a regrouping; plastic deformation and boundary diffusion play a leading role during this stage, while the contribution of bulk diffusion is negligible. Boundary diffusion approximation is justified in the temperature interval where the boundary mass transfer is dominant over the one in the bulk of a crystal. The intermediate stage is the most complicated one. The final stage is characterized by the presence of isolated pores in a sample. Kinetics of a pore system in this stage is controlled by mass transfer along grain boundaries and dislocation lines (if available) at lower temperatures and by bulk mass transfer mechanisms at higher temperatures.

Remaining after the final stage, isolated pores influence physical-mechanical properties of a product undesirably in the majority of cases. Arising after coagulation, large pores are usually especially harmful and their removal is especially difficult. Electro-physical technologies, such as electric contact, hf, inductive heating, enable overcoming of such difficulties in a series of cases [2].

The author had formulated the General Thermodynamic Theory of Sintering in 1998 [3, 4]. In the present paper the author describes this theory and applies it to the problem of swelling.

## 2. BULK HEATING

It was predicted theoretically that the evolution of heat in the bulk of a sample (e.g., Joule heat) leads to specific driving forces of diffusional processes [5, 6]. The sources of these driving forces are divergent temperature gradients, arising as a result of the bulk heating of a sample (as the heat, evolving in the bulk of a sample has to be extracted out of a sample, the temperature, $T$, appears to be higher in the bulk of a sample compared to the temperature of the surface of a sample) [2, 5, 6]. Fig. 1 shows temperature gradients, arising during bulk heating process when average temperature of a sample stays unchanged.

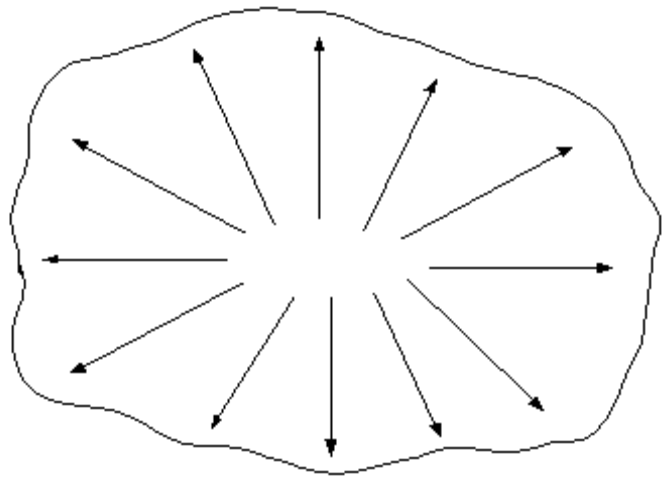

*Fig. 1. Heat flow out of a sample during bulk heating process (the sample averaged temperature is assumed to be fixed).*

Fixed sample averaged temperature assumes that the amount of heat evolving in a sample is equal to the amount of heat being extracted out of a sample.

### 2.1. Driving Forces

The temperature gradient causes the thermal diffusion flux of vacancies,

$$\mathbf{J} = -k_T (D/\langle T \rangle) \mathrm{grad} T, \qquad (1)$$

where $k_T$ is the thermal diffusion ratio, $D = D(\langle T \rangle)$ is the diffusion coefficient, and $\langle T \rangle$ is the temperature of a sample averaged over its volume.

Eq. (1) is a linear approximation, which is applicable when the temperature gradient is smooth enough and small enough, which is the case.

Eq. (1) determines the additional number of vacancies leaving unit volume of a material per unit time due to the regarded driving forces:

$$n_t = \mathrm{div}\mathbf{J} = -Dk_T \Delta T/\langle T\rangle = Dk_T(q - cT_t)/\kappa\langle T\rangle, \tag{2}$$

where $q$ is the heat production per unit volume of a material per unit time (in the case of ohmic heating this is the Joule heat, $q = \sigma E^2$, where $\sigma$ is the specific electric conductivity, $E$ is the intensity of the effective electric field), $c$ is the heat capacity (per unit volume), $T_t = (\partial T/\partial t)$ is the heating (cooling) rate, and $\kappa$ is the thermal conductivity coefficient.

To obtain Eq. (2), the equation of the heat flow,

$$\kappa\Delta T = (cT_t - q), \tag{3}$$

was used [2].

The thermal diffusion ratio, $k_T$, is determined by the formula [2],

$$k_T = N_e(u - u_m)/kT, \tag{4}$$

where $N_e$ is the mean equilibrium number of vacancies per unit volume of a material at temperature $T$, $u$ is the vacancy formation enthalpy for the given process of diffusional transport (e.g., in the case of the transport of vacancies from the cores of dislocations to the plane surface of a sample, $u = u_d$ is the enthalpy of vacancy formation on dislocations), $u_m$ is the enthalpy of vacancy migration, and $k$ is Boltzmann constant.

The contribution of specific driving forces, arising from bulk heating is the most important at the isothermal regime, when $T_t = 0$, and at large enough values of the power, $q$, being introduced into the bulk of a material. Such conditions are readily fulfilled during sintering of thin or miniature samples, when heat emission from the surface contributes essentially to the heat balance. This contribution may be decisive during ohmic sintering of metallic samples thinner than 1 mm. In this case electric field intensities, $E$, of the order of 1 V/cm and larger may be applied [7]. Under these conditions healing of large pores differs very significantly from that during conventional (furnace) sintering. This follows from the fact that the contribution of the specific driving forces accordingly to Eq. (2) does not depend on the nature of defects, and, in particular, on the pore size. However, the Laplace forces increase with decreasing pore size [1, 2], thus reducing the relative contribution of bulk heating driving forces to the process of pore healing.

In some cases it is worthwhile to intensify surface heat emission (to cool the surface of a sample). In this case the value of $q$ can be increased and the time required to complete the diffusional process is reduced concomitantly. As a result, the energy required to complete the process is smaller than that during conventional sintering. One ought to take into account the advantages of more compact and energy-saving equipment in the case of ohmic or hf sintering as compared to using a traditional furnace.

**2.2. Initial Stages of Bulk Heating Sintering**

The very first stages of electric sintering are essentially non-stationary ones. Evolution of Joule heat during the very first moments of electric sintering leads to sharp temperature differences. Such an essentially non-stationary regime of heat and mass transfer can only be described by an essentially non-linear theory. A similar situation arises when sintering is caused by electric discharge [8]. Here we shall consider early, but not the very first, stages of electric sintering [9]. The temperature gradients are considered to be small, and linear approximation is assumed to be applicable. Boundary diffusion is regarded to be the main mechanism of mass transfer. In this case the number of atoms, leaving one contact per unit time on one particle, $N_{at}$, is described by the integral of $(\mathbf{J}_a,\mathbf{n})$ along the closed path, surrounding the contact (here $\mathbf{J}_a$ is the atomic flux, and $\mathbf{n}$ is the unit vector, perpendicular to the path). The line integral around the closed path can be transformed to the surface integral of div$\mathbf{J}_a$ (the integral is extended over the surface of a contact). The processes of mass transfer during electric sintering are much faster than those, arising due to surface tension forces in the course of conventional sintering. This means that the influence of surface tension forces is often negligible during electric sintering. In this case the surface flux of the atoms, $\mathbf{J}_a$, can be expressed as

$$\mathbf{J}_a = -D_b(k_{Tb}/\langle T \rangle)\mathrm{grad}T, \tag{5}$$

where $D_b$ is the boundary diffusion coefficient, and $k_{Tb}$ is a boundary thermal diffusion ratio for atom.

Let us the consider Joule heating under isothermal conditions. When a component of the temperature gradient, perpendicular to the boundary is not taken into account, the number of atoms, leaving the contact area on a single grain per unit time is given by the following expression:

$$N_a = D_b \sigma E^2 s k_{Tb}/\kappa \langle T \rangle, \tag{6}$$

where $\sigma$, $E$, $\kappa$ are local electric conductivity, electric field, thermal conductivity on the contact, and $s$ is the area of the contact.

If perpendicular to the boundary component of the temperature gradient is taken into account, the right-hand part of Eq. (6) acquires an inessential factor of the order of unity.

As the current passes two contacts in each particle, we have for the rate of the process:

$$-V_t = 2v_a N_{at} = 2D_b \sigma E^2 k_{Tb} v_a s(t)/\kappa \langle T \rangle, \tag{7}$$

where $V$ is the volume of a sample per number of constituting particles, and $v_a$ is the atomic volume.

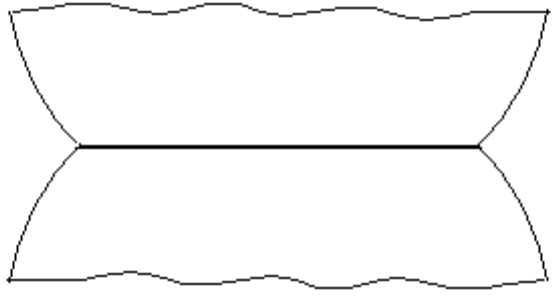

*Fig. 2. Model of a contact for the early stages of electric contact sintering.*

Let us consider now a model of identical particles of an initial radius $R_0$. To simplify the calculations let us assume that after pressing and during the considered early period of sintering each particle is a cubically symmetric figure, formed by a sphere of a radius $R(t)$, of which 6 segments are cut off by planes (see Fig. 2). We assume that at the early stages of sintering $s(t)$ is essentially smaller than $R^2$, although in real powder systems $s(t)$ and $R^2$ are often of the same order of magnitude. This model is justified, because the contribution of the regular diffusional processes to the sintering and to the shape of the contact is negligible. For the sake of simplicity of the model the anisotropy, arising due to the electric current is not taken into account.

In the first approximation on $[s(t)/R^2]$ small parameter we get that

$$V = 8R_0^3 - (22/\pi)R_0 s. \tag{8}$$

Eq. (7) and (8) yield

$$(V_0 - V)_t = A(V_0 - V), \; A = \pi D_b \sigma E^2 k_{Tb} v_a / 11 R_0 \kappa \langle T \rangle, \; V_0 = 8R_0^3, \tag{9}$$

where $V_0$ is the initial volume of a sample per one particle.

Eq. (9) yields

$$V(t) = V_0 - (22/\pi) R_0 s(0) \exp At. \tag{10}$$

As follows from Eq. (10), the shrinkage kinetics equation has a form,

$$[V(0) - V(t)]/V(0) \equiv \delta V(t)/V(0) = [33 s(0)/2\pi^2 R_0^2][(\exp At) - 1]. \tag{11}$$

It is worthwhile to note that Eq. (11) is applicable as far as its right-hand part is essentially smaller than the maximum possible value of shrinkage, which is assumed to be essentially smaller than unity. Fig. 3 shows the dependence of a relative shrinkage, $\delta V(t)/V(0)$, on time, $t$, for the early stages of electric contact sintering.

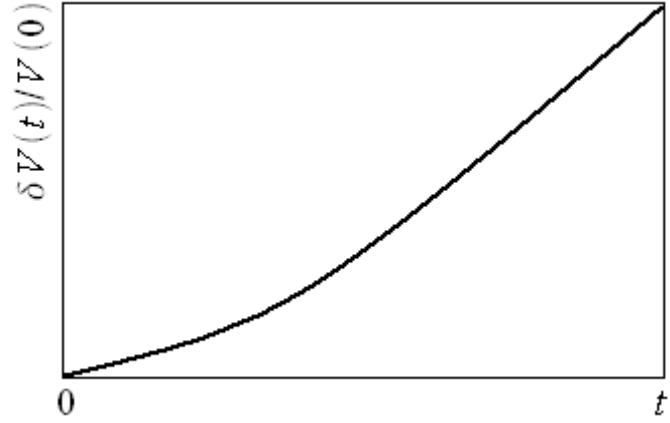

*Fig. 3. Sample relative shrinkage dependence on time (early stages of electric contact sintering)*

Eq. (11) was obtained under the assumption that parameter *A* has not changed during the regarded process. This assumption is valid when heat evolution, *q*, and also the effective electric field, *E*, do not change during this process. During initial stages of electric sintering and when $s(t) \ll R_0^2$, the contribution of the contacts into electric resistance of a sample is the main one, and therefore the effective electric field in the contacts remains practically unchanged during the regarded process. Changes in the effective electric field, *E*, during the regarded initial stages of electric sintering could be taken into account by retaining the second and higher powers of the small parameter, $[s(t)/R_0^2]$, when performing calculations.

**2.3. Final Stages of Bulk Heating Sintering**

Let us consider processes determined by diffusion of vacancies through the bulk of a material. Let us assume that our sample contains several ensembles of sources (sinks) of vacancies, which are identical in each given ensemble. Each *i*-th ensemble is characterized by its own equilibrium value of vacancy concentration, $N_{ei}$, an enthalpy of the vacancy formation on any source (sink) of the ensemble, $u_i$, and a relaxation time of the number of vacancies to their equilibrium value, $\tau_i$. In this case in a stationary regime the continuity equation for vacancies can be written as follows:

$$\text{div}\mathbf{J} + \Sigma_i[(N - N_{ei})/\tau_i] = 0, \qquad (12)$$

where *N* is the vacancy concentration, and the vacancy flux, **J**, is as follows:

$$\mathbf{J} = -D\,\text{grad}N + k_{Ta}(D/\langle T \rangle)\text{grad}T, \qquad (13)$$

where $k_{Ta}$ is the thermal diffusion ratio for the atoms.

When $(u_i - u_j) \ll kT$, and a linear approximation of the transport theory is applicable, the equation for the excess vacancies concentration, $n_i = N - N_{ei}$, is as follows:

$$[\Delta - \Sigma_j(1/D\tau_j)]n_i = (\sigma E^2 k_T/\kappa\langle T\rangle) + N_e\Sigma_j[(u_j - u_i)/D\tau_j k\langle T\rangle], \tag{14}$$

where the thermal diffusion ratio for the vacancies,

$$k_T = (N_e u/k\langle T\rangle) - k_{Ta}. \tag{15}$$

In a simple atomic theory [10] Eq. (4) takes place, which may be written here as follows:

$$k_T = N_e(u - u_m)/k\langle T\rangle. \tag{16}$$

The quantity $(D\tau_i)^{1/2}$ has the meaning of a mean free path of the excess vacancy, $\langle l\rangle$, in the bulk of a material related to the $i$-th type sinks.

The rate of the diffusional changes of the $i$-th type sources (sinks) is proportional to $n_i$. For example, for pores, the rate of changes in the pore volume of pores of the $i$-th type, $(v_i)_t$, is described by the following relationship:

$$(v_i)_t = n_i/N_0 n_{pi} \tau_i, \tag{17}$$

where $N_0$ is the number of the atoms in the unit volume of a material, $n_{pi}$ is the number of pores of the $i$-th type in the unit volume.

The quantity $n_i$ is proportional to the right-hand part of Eq. (14), $E$, and other parameters, which determine the value and sign of it. In the bulk of a material the Laplacean in the left-hand part of the Eq. (14) can be neglected and thus $n_i$ can be determined without solving the differential equation of the second order.

For a pore of the $i$-th type, when the pressure inside the pore and external pressure are neglected [1, 2], we have:

$$u_i = u_0 - (2\gamma/r_i N_0), \tag{18}$$

where $u_0$ is the enthalpy of the vacancy formation on the plane surface, $\gamma$ is the specific surface energy, and $r_i$ is the pore radius.

The enthalpy of vacancy formation on smaller sized pores is lower [1, 2]. According to this at $E = 0$ Eqs. (14) and (18) describe the process of coagulation of small pores into large ones [2 – 4]. It follows from Eq. (14) that the pores of ensembles, which right-hand part of Eq. (14) is positive, shrink, and these are the ensembles of smaller pores [2]. The condition of pore healing is as follows:

$$r_i < r^* \equiv \Sigma_j\{1/\tau_j\{[\Sigma_k'(1/r_k\tau_k)] - (\sigma E^2 k_T N_0 kD/2\gamma\kappa N_e)\}\}, \tag{19}$$

where $\Sigma_k'$ denotes a summation over the ensembles of pores only. Only pores with a radius smaller than $r^*$ are being healed.

At $k_T > 0$ $r^*$ increases monotonically with the increase in $E$, going asymptotically to infinity when $E \to E^*$,

$$E^* \equiv [2\gamma\kappa N_e\Sigma_k'(1/r_k\tau_k)/kD\sigma k_T N_0]^{1/2}. \tag{20}$$

At $E \geq E^*$, according to Eq. (19), no pore can grow.

Eq. (19) can be represented in another form:

$$E > E_i \equiv \{(2\gamma\kappa N_e/kD\sigma k_T N_0) | [\Sigma_k'(r_i - r_k)/r_i r_k \tau_k] - (1/r_i)\Sigma_k''(1/\tau_k) | \}^{1/2}, \quad (21)$$

where $\Sigma_k''$ denotes a summation over all ensembles of sinks and sources without the ensemble of pores.

At $k_T > 0$ Eq. (21) determines the field value, $E_i$, which is sufficient to cause shrinking of the pores of the $i$-th ensemble.

2.3.1. The Role of the Mobile and Immobile Dislocations

This role was considered in [11]. Eq. (14) shows that the rates of diffusional processes are determined essentially by the differences in vacancy formation enthalpies on different types of sources (sinks) and by mean free paths of excess vacancies in respect to different ensembles of sources (sinks), $(D\tau_i)^{1/2}$. The greater the number of sources (sinks), present in a crystal, the smaller the mean free path of an excess vacancy and the higher the rate of diffusional processes [12].

Dislocations are typical sources (sinks) of vacancies in a crystal. However, when the density of dislocations increases, many segments of dislocations, blocked by intersections with another dislocation lines, are formed (see Fig. 4). Such segments cannot serve as sources (sinks) for a considerable while, as their function as sources (sinks) results in bending of the segments between the points of the intersections, and therefore to the increase in the enthalpy of vacancy formation on the regarded segment of the dislocation (impurity atoms play, obviously, an important role in immobilizing of intersection points). That is why only thermally mobile dislocations should be taken into account as possible sources (sinks) of vacancies. Now let us calculate the density of thermally mobile dislocations [11].

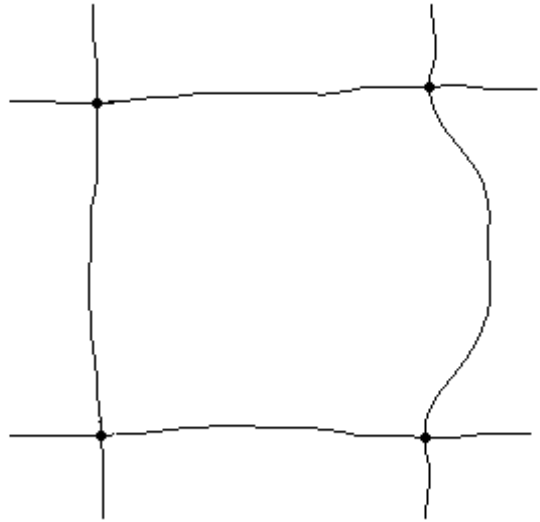

*Fig. 4. Mobile and immobile dislocations.*

Let us consider randomly distributed linear dislocations, parallel to the Descartes' coordinates. Let the densities of dislocations of each of the three directions be the same and equal to 1/3 of the total density, $n_d$. Let us approximate a line of a dislocation by a square cylinder with a base side, $b$, ($b$ is of the order of magnitude of the Burgers vector), and a grain (in the case of a single crystal - the sample) - by a cube with a rib, $L$. Let us regard some dislocation and some intersection point. Obviously, the next intersection point can be found at the distance between $x$ and $x + dx$ with the probability,

$$w(x)dx = [1 - (xb/L^2)]^{2n_dLL/3}(2n_d bL^2 dx/3L^2). \qquad (22)$$

Taking into account that $2n_dL^2 \gg 3$, one can calculate the number of segments of the length between $x$ and $x + dx$: $(2/3)n_d bL[\exp(-2/3)n_d bx](2/3)n_d b dx$.

Let us assume that the segments shorter than some characteristic length, $l$, ($x \leq l$) cannot serve as sources (sinks) in the regarded diffusional process. The total length of such segments on one dislocation can be calculated. For this the number of segments of the length between $x$ and $x + dx$ should be multiplied by $x$ and integrated from 0 to $l$. The result is: $L\{1 - [1 + (2/3)n_d bl]\exp[-(2/3)n_d bl]\}$.

The ratio of this length to the length of the dislocation, $L$, is a relative density of thermally immobile dislocations. Therefore for the density of thermally mobile dislocations, $n_{dt}$, we have:

$$n_{dt} = n_d[1 + (2/3)n_d bl]\exp[-(2/3)n_d bl]. \qquad (23)$$

Obviously, $n_{dt} \approx n_d$ at $2n_d bl \ll 3$; then, with the increase in $n_d bl$, $n_{dt}$ achieves a maximum at $n_d = n_m$, and after that $n_{dt}$ decays exponentially with further increase in $n_d bl$ (see Fig. 5). The value of the maximum, $n_{tm} = 1.26/bl$, is achieved at $n_d = n_m = 2.43/bl$.

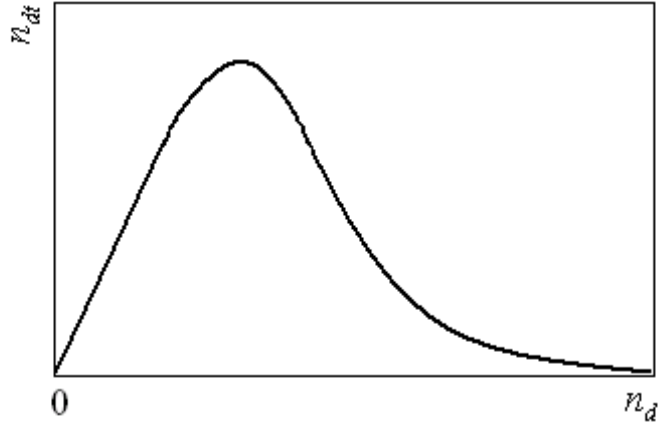

*Fig. 5. Thermally mobile dislocations density, $n_{dt}$, as a function of general dislocation density, $n_d$.*

The density of thermally mobile dislocations, arising in the process of annealing of deformed bcc iron, containing a small amount of impurities, was measured (see [13]). At $n_d = 3 \times 10^{11}$ cm$^{-2}$, the mobile dislocation density, $n_{dt}$, was $10^{11}$ cm$^{-2}$. With these data values Eq. (23) yields $bl = 1.145 \times 10^{-11}$ cm$^2$. At $b = 3 \times 10^{-8}$ cm this corresponds to $l = 3.8 \times 10^{-4}$ cm. The maximum density of thermally mobile dislocations, $n_{tm}$, is $1.1 \times 10^{11}$ cm$^{-2}$ at $n_m = 2.22 \times 10^{11}$ cm$^{-2}$. It is worthwhile to note that in the regarded case the maximum is a rather pronounced one. At $10^{-11}n_d = 1$, 2.22, 3, and 6, $10^{-10}n_{dt}$ were 8.2, 11, 10, and 3.4.

So, when the intersection points of the dislocation lines are pinned by, e.g., impurity atoms, an optimal dislocation density exists at which the density of thermally mobile dislocations is of a maximum value and the rates of diffusional processes also have maximum values. The diffusional stage of the process of sintering is one of such processes.

2.3.2. Calculation of Relaxation Parameters.
In order to calculate the relaxation parameters, let us consider a case when there is only one type of sources (sinks) of vacancies in a crystal and there is a macroscopically homogeneous excess of the vacancy concentration in a sample. It is assumed that the vacancy concentration in the vicinity of the sources (sinks) has an equilibrium value. The coordinate dependence of the vacancy concentration around the given source (sink), in the domain of its influence (drainage basin), is determined by the stationary distribution, that is, by the Laplace equation. The relaxation parameter, $1/D\tau$, will be further calculated for some typical sorts of sources (sinks) of vacancies (see, e.g., [12]).

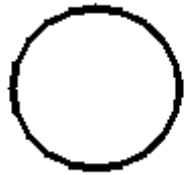

*Fig. 6. Model of a grain and its boundary as a sink.*

***Grain boundaries***. Let us approximate a grain by a ball (see Fig. 6) of a diameter $d$, and a microscopic dependence of the vacancy concentration in the vicinity of a grain boundary, $N$, on the distance from the center of a grain, $r$, by the following equation:

$$N = N_e + (a/r) - (2a/d), \qquad (24)$$

where $a$ is some constant.

Then the vacancy flux near the grain boundary is $4aD/d^2$, and the number of vacancies, coming to a grain boundary per unit time, and calculated per unit volume, is $24aD/d^3$, which, on the other hand, may be written as $(\langle N \rangle - N_e)/\tau = a/d\tau$. From this follows that $1/D\tau = 24/d^2$ for grain boundaries, which means that the mean free path of the excess vacancy inside a grain with respect to the grain boundary is about $l_{gb} = 0.2d$.

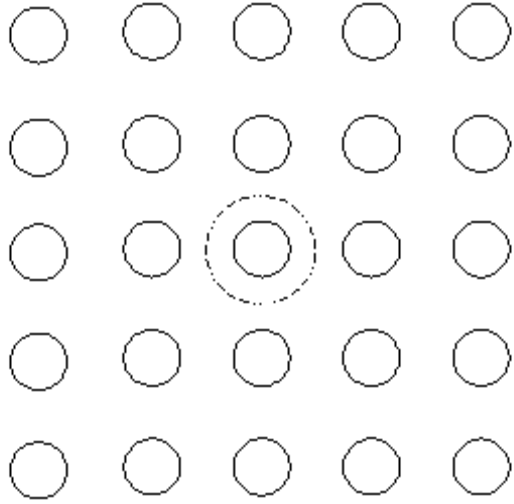

*Fig. 7. Model of a pore as a sink. Drainage basin is plotted by a dotted line.*

***Pores***. Let us consider an ensemble of identical homogeneously distributed spherical pores (see Fig. 7) of a radius, $r_0$, and a pore concentration, $n_p$. The porosity is assumed to be small. This allows us to approximate a drainage basin of a pore, a volume of influence of each pore (a volume from which a given pore attracts excess vacancies), by a sphere of a radius $0.62 n_p^{-1/3}$, and to approximate a microscopic dependence of the vacancy concentration on the distance from the center of the pore, $r$, as

$$N = N_e + (a/r_0) - (a/r). \qquad (25)$$

The vacancy flux near the surface of a pore is $aD/r_0^2$, and the number of vacancies, coming to the surface of pores per unit time, and calculated per unit volume, is $4\pi a D n_p$, which, on the other hand, may be written as $(\langle N \rangle - N_e)/\tau = a/r_0\tau$. From this follows that $1/D\tau = 4\pi n_p r_0$ for the pores.

Let us consider a case when there are spherical pores of different radii, $r_k$, in the sample, but no other defects. Then we have

$$\Sigma_k(1/\tau_k) = 4\pi D \langle r_p \rangle \langle n_p \rangle, \ \Sigma_k(1/r_k\tau_k) = 4\pi D \langle n_p \rangle, \qquad (26)$$

where $\langle r_p \rangle$ is the average radius of the pores and $\langle n_p \rangle$ is the total number of pores in the unit volume of a sample.

***Dislocations***. Speaking about dislocations, one should take into account that vacancies are generated and annihilated by dislocation jogs. When the averaged distance between the jogs, $l_j \geq n_{dt}^{-1/2}$ (low jog density), the jogs could be regarded as a limiting case of pores, where the size of the Burgers vector, $b$, corresponds to the radius of a pore, $r_0$, and so the following equation can be applied:

$$1/D\tau = 4\pi b n_{dt}/l_j. \qquad (27)$$

Here an interaction between a vacancy and a stress field of dislocations is not taken into account. This interaction can be neglected when its energy is much smaller than $kT$, which is the case at ambient temperature and higher ones. At $n_{dt} = 10^{10}$ cm$^{-2}$, $l_j = 10^{-5}$ cm, $b = 3\times 10^{-8}$ cm, Eq. (27) yields $1/D\tau = 4.19\times 10^8$ cm$^{-2}$.

When the jog density is very high (high jog density), $l_j \ll n_{dt}^{-1/2}$, almost all the volume of a sample can be approximated by cylinders (see Fig. 8), surrounding parallel, uniformly situated dislocations. The radius of the cylinder is about $0.564 n_{dt}^{-1/2}$. The concentration of vacancies in a separate cylinder can be approximated as

$$N = N_e - a\ln(\rho/l_j), \tag{28}$$

where $\rho$ is the distance from the dislocation axis.

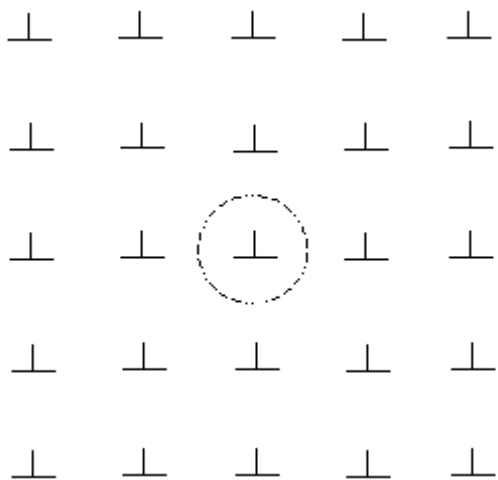

Fig. 8. Model of a dislocation as a sink for high jog density case. Drainage basin is plotted by a dotted line.

It is accepted in Eq. (28) that $N = N_e$ when $\rho \approx l_j$. The vacancy flux in the vicinity of the cylinder, where $\rho \approx l_j$, is $-aD/l_j$. The number of vacancies, crossing the surface of the cylinder, $\rho \approx l_j$, in the unit time, and calculated per unit volume, is $2\pi a D n_{dt}$. On the other hand this number is equal to the number of vacancies, annihilating on dislocations per unit time, $(\langle N \rangle - N_e)/\tau = -(a/2\tau)\ln(8.54 l_j^2 n_{dt})$. From this follows that $1/D\tau = -4\pi n_{dt}/\ln(8.54 l_j^2 n_{dt})$. At $8.54 l_j^2 n_{dt} = 10^{-3}$ and $n_{dt} = 10^{10}$ cm$^{-2}$ we have $1/D\tau = 1.82\times 10^{10}$ cm$^{-2}$.

It is worthwhile to note that relaxation parameter, $1/D\tau$, in the regarded case is not very sensitive one to the value of $l_j$ (the dependence is a logarithmic one).

***The role of the quality of the surface***. In cases when the process depends upon mass transfer to or from surfaces (for example, pore surfaces), its rate is determined in many aspects by both the structure and quality of the surface [14]. Usually a vacancy concentration in a thin layer near the surface is assumed to be equal to the equilibrium value with respect to the vacancy formation in the given spot of the surface. This is valid only in the case when the time interval during which a vacancy is formed on the surface is considerably smaller than that during which a vacancy exists in the above layer. This condition is fulfilled usually in perfect crystals, where diffusional currents are small. Diffusional currents in real crystals may be so intense that equilibrium cannot be achieved. The rate of approach of a vacancy concentration to the equilibrium one near the surface depends on the density of sources (sinks) of vacancies on the surface (surface defects). When the density of these defects is sufficiently high, the above rate is considerable, and the mass transfer process proceeds quickly. Otherwise, the rate of the surface motion is slow.

Calculations show [14] that the rate of the change of the isolated pore of the radius $r$ has a factor $v/[D\langle l \rangle + (D + v\langle l \rangle)r]$, where $v$ is a parameter, which characterizes the ability of the ensemble of surface sources to supply deficient vacancies and $\langle l \rangle$ is the mean free path of the excess vacancy in the bulk of a material.

2.3.3. Theoretical Conclusions, Concerning Final Stages of Sintering

On the basis of results, discussed above, equations, describing the kinetics of pores of the $i$-th ensemble were derived. In a general non-stationary case it has a form [3]:

$$(dr_i/dt) = -(D/N_0 r_i)[(6/\pi d^2) + (bn_{dt}/l_j) + \langle n_p \rangle \langle r_p \rangle]^{-1}\{[k_T(\sigma E^2 - cT_t)/4\pi\kappa\langle T \rangle] +$$

$$+ (N_e/k\langle T \rangle)[(6/\pi d^2)(u_{gb} - u_0) + (bn_{dt}/l_j)(u_d - u_0)] + (2\gamma N_e/k\langle T \rangle N_0 r_i) \times$$

$$\times [(6/\pi d^2) + (bn_{dt}/l_j) + \langle n_p \rangle (\langle r_p \rangle - r_i)]\}, \tag{29}$$

where $r_i$ is the radius of a pore, belonging to the $i$-th ensemble, $u_{gb}$ and $u_d$ are the enthalpies of vacancy formation on grain boundaries and dislocations, respectively.

Kinetics of the averaged pore radius is described by the following equation [3]:

$$(d\langle r_p \rangle/dt) = -(D/N_0)[(6/\pi d^2) + (bn_{dt}/l_j) + \langle n_p \rangle \langle r_p \rangle]^{-1}\{[k_T(\sigma E^2 - cT_t)/4\pi\kappa\langle T \rangle]\langle 1/r_p \rangle +$$

$$+ (N_e/k\langle T \rangle)[(6/\pi d^2)(u_{gb} - u_0) + (bn_{dt}/l_j)(u_d - u_0) - (2\gamma N_e/k\langle T \rangle N_0)\langle n_p \rangle]\langle 1/r_p \rangle +$$

$$+ (2\gamma N_e/k\langle T \rangle N_0)[(6/\pi d^2) + (bn_{dt}/l_j) + \langle n_p \rangle \langle r_p \rangle]\langle 1/r_p^2 \rangle\}. \tag{30}$$

On the basis of the described theoretical results, we may conclude that the rate of the given diffusional process is determined by the following factors:

*Heating method*. Bulk heating effectively accelerates mass transport processes when the production of heat in the bulk of a material, $q$, is high enough. This occurs when small or thin products, with a thickness of the order of 1 mm or thinner, are sintered.

*The structure of crystalline defects of sintered powders*. The higher the concentration of defects, the more intense the mass transport processes are, as a rule. But the dependence of the process rate on the dislocation density is not monotonic. It has a maximum at some dislocation density, whose typical value is about $10^{11}$ cm$^{-2}$.

*The ability of both the bulk and surface vacancy sources to emit or absorb vacancies*. As healing of large pores (larger than 1 μ in diameter) is enhanced due to bulk heating, the process of coagulation of small pores into large ones is impeded (in contrast to conventional sintering). Hence, the conclusion may be drawn that the samples sintered by bulk heating will contain a larger number of small pores and smaller number of large pores than found in the regular furnace sintered samples. Such a more uniform pore structure, consisting mostly of small pores, tends to improve mechanical and other physical properties of products.

A comparative experimental investigation of sintering of iron powder by means of electric contact, inductive, and conventional furnace sintering was performed [15]. The results showed that electric contact sintering allows achieving of the same level of mechanical properties during an essentially smaller period of treatment compared to conventional furnace sintering.

**3. Cycling and Pulsing Regimes**

During cycling around a temperature interval of the first order phase transformation intensification of diffusional mass transfer processes may be observed in some cases (see, e.g., [3]). Such an influence takes place only when the phase transformation is of a martensitic type [16], as in this case every separate phase transformation is accompanied by some plastic deformation during which excess vacancies are formed. Formation of excess vacancies causes changes in sources (sinks) of vacancies, which is a mass transfer by itself, and this enhances the process of sintering. Vacancies are formed on locations where the enthalpy of vacancy formation is smaller. Then the excess vacancies annihilate on locations where the enthalpy of the vacancy formation is larger. This process contributes to a further development of process of diffusional sintering. Other processes of mass transfer, typical for non-isothermal electric sintering, take place concomitantly and were investigated in [3].

An equation, which determines time and coordinate dependence of temperature, is the heat flow equation, Eq. (3). At cycling or pulsing regimes, the heat production, $q = \sigma E^2$, varies periodically. Let us separate all the sources (sinks) of vacancies, existing in a sample, into two grades. Let the first grade be the one, which contributes to the sintering process directly, and the second one does not contribute to the sintering process directly [e.g., dislocations; they do not contribute immediately to the process of sintering (consolidation of a material) during their activity as sources (sinks)]. As was shown [17], all sources (sinks), which are described by parameters $\tau_i$ and $N_{ei}$, can be replaced by one effective type of source (sink) with

$$(1/\tau) = \Sigma_i (1/\tau_i) \text{ and } N_e = \tau \Sigma_i (N_{ei}/\tau_i). \tag{31}$$

The dependence of the vacancy concentration on time and coordinates during the process of sintering is determined by the continuity equation:

$$\text{div}\mathbf{J} + [(N - N_{e1})/\tau_1] + [(N - N_{e2})/\tau] + (\partial N/\partial t) = N_{tv}, \tag{32}$$

where $N_{e1}$ and $\tau_1$ refers to the first grade sources (sinks), $N_{e2}$ and $\tau_2$ refers to the second grade ones, and $N_{tv} = (dN_v/dt)$ is the production of vacancies per unit time in the unit volume of the sample, caused by direct and reverse martensitic phase transitions. The vacancy flux, $\mathbf{J}$, is determined by Eq. (13).

Substituting Eq. (13) into Eq. (32), and taking into consideration Eq. (3), we obtain equation, which determines time and coordinate dependence of the vacancy concentration:

$$\{D\Delta - [(\tau_1 + \tau_2)/\tau_1 \tau_2]\}(N - N_{e1}) = (\partial N/\partial t) +$$

$$+ (Dk_T/\kappa\langle T\rangle)(cT_t - q) + [(N_{e1} - N_{e2})/\tau_2] - N_{tv}. \tag{33}$$

Changes occurring during time intervals considerably larger than the cycling period are of primary interest. So let us time-average Eq. (33) over the cycle period. Let us consider only the bulk of a material, where macroscopic inhomogeneity of the vacancy distribution is inessential. This inhomogeneity is essential only in a surface layer of the thickness of about $[D\tau_1\tau_2/(\tau_1 + \tau_2)]^{1/2}$. When time-averaged temperature and vacancy concentration remain unchanged (quasi-stationary regime), we obtain the following relationship, valid in the bulk of a sample:

$$\langle[(N - N_{e1})/\tau_1]\rangle_t = \langle[(N_{e2} - N)/\tau_2]\rangle_t + \langle qDk_T/\kappa\langle T\rangle\rangle_t + \langle dN_v/dt\rangle_t. \qquad (34)$$

The left-hand part of Eq. (34) refers to first grade processes, that is, to the rate of the sintering process. The right-hand part of Eq. (34) represents the action of different sources and driving forces of diffusional processes. The first term represents conventional sources and driving forces of furnace sintering. The second term describes the contribution of divergent thermal diffusion fluxes, caused by the inhomogeneity of the temperature gradient in the bulk-heated sample. The third term describes a contribution of excess vacancies, formed by plastic deformation during direct and reverse martensitic transformations during cycling. The larger the third term the greater the enhancement of the sintering process is. From this follows, that the cycling and pulsing regimes are more effective at higher frequencies (shorter periods).

It is worthwhile to regard separately two following cases:
**(i)** First of all let us consider a case when $N_{e1} < N < N_{e2}$. In this case sources of a second grade emit vacancies, which annihilate on the sinks of a first grade. In the early stages of sintering this leads in some cases to the decrease in the curvature of grains and thus the process of sintering progresses. As in the regarded case $N < N_{e2}$, the contributions of the first and the third terms in the right-hand part of the Eq. (34) are positive. This means that the rate of sintering at the early stages of the process increases monotonically with the increase in the frequency of cycling.

The final stages of sintering consist in the healing of pores. So to enhance the process of sintering we have to create conditions when the source of a first grade (e.g., pores) emits vacancies. But in the regarded case first grade sinks absorb vacancies. Thus the sintering process (in its final stages) is impeded, and the higher the cycling frequency, the slower the rate of the final stages of sintering is.

**(ii)** Now let us regard a case when $N_{e2} < N < N_{e1}$. In this case the first grade sources emit vacancies, which annihilate on second grade sinks. That is why in the regarded case increase in the excess vacancy concentration impedes the bulk vacancy diffusion mechanism of the activity of the first grade sources, thus diminishing a negative contribution of this mechanism to the early stages of sintering. So, in the early stages, cycling and pulsing enhances sintering, and the higher the frequency, the more essential the enhancement is. As in the regarded case the emission of vacancies by the first grade sources is slowed down when excess vacancy concentration is increased, the rate of the final stages of sintering (the healing of the pores) is slowed down also, and the higher the cycling frequency the lower the rate of the final stages of sintering is.

## 4. THE PROBLEM OF SWELLING

The problem of swelling was discussed in [17, 19]. Here we shall discuss it in more detail [4]. Penetrating irradiation dislocates some atoms of the lattice into interstitial positions, leaving behind vacancies. Coagulation of vacancies leads to pore formation. As a result of this process the material swells. Swelling materials loose their mechanical and other properties and they do not function properly. Let us consider the process of swelling analytically. Let us assume that in the unit volume of a sample during a unit time interval $N_{ti}$ dislocated atoms and $N_{tv}$ vacancies are created. The dislocated atoms and vacancies annihilate on sinks (grain boundaries, dislocations and pores) as was described above [Eq. (12)]. Some of them are being driven out of the sample by the divergent temperature gradients as was shown above [Eq. (2)]. When the quantity of vacancies, driven out of the sample by divergent temperature gradients, achieves the quantity of vacancies, created in the sample by penetrating radiation, that is when

$$q \geq N_{tv} k \kappa \langle T \rangle^2 / [D_v N_{ev}(u_v - u_{nv})], \qquad (35)$$

the swelling is stopped [17, 19]. Eqs. (2) and (16) were used to obtain Eq. (35).

Now let us take a more general look at the problem of swelling prevention. Let us write the balance for the stationary regime:

$$N_{ti} = (N_i/\tau_{gbi}) + (N_i/\tau_{di}) + (N_i/\tau_{pi}) + (D_i q k_{Ti}/\kappa \langle T \rangle), \; N_{tv} = [(N_v - N_{ev})/\tau_{gbv}] +$$

$$+ [(N_v - N_{ev})/\tau_{dv}] + [(N_v - N_{ev})/\tau_{pv}] + (D_v q k_{Tv}/\kappa \langle T \rangle). \qquad (36)$$

Here $N_i$ is the concentration of dislocated atoms, $N_v$ is the vacancy concentration. The first line refers to the balance of dislocated atoms; the second one refers to the balance of vacancies. The equilibrium value of interstitial atoms was neglected in Eq. (36), as it is very small. The rate of annihilation of excess vacancies on sinks is assumed to be proportional to the excess vacancy concentration, $(N_v - N_{ev})$. The relaxation parameters used in Eq. (36) were calculated in Section 2.3.2. The diffusion coefficients for dislocated atoms and vacancies, $D_i$ and $D_v$ respectively, are different. We assume that the concentration of dislocated atoms and vacancies is not too high, so the process of mutual annihilation of dislocated atoms and vacancies is neglected in Eq. (36). This process could be described by additional terms in both lines of Eq. (36), containing a product of both $N_i$ and $N_v$.

Eq. (36) yields:

$$N_i = [N_{ti} - (D_i q k_{Ti}/\kappa \langle T \rangle)]/[(1/\tau_{gbi}) + (1/\tau_{di}) + (1/\tau_{pi})],$$

$$N_v - N_{ev} = [N_{tv} - (D_v q k_{Tv}/\kappa \langle T \rangle)]/[(1/\tau_{gbv}) + (1/\tau_{dv}) + (1/\tau_{pv})]. \qquad (37)$$

The condition for a pore not to grow is that the number of atoms, coming to the pore prevails upon the number of coming vacancies. Using Eq. (37), one can write this condition in the form:

$$[N_{ti} - N_{tv} + (D_v q k_{Tv}/\kappa \langle T \rangle) - (D_i q k_{Ti}/\kappa \langle T \rangle)]/[1 + (\tau_p/\tau_{gb}) + (\tau_p/\tau_d)] \geq 0. \qquad (38)$$

It is worthwhile to note that ($\tau_p/\tau_{gb}$) and ($\tau_p/\tau_d$) do not contain the diffusion coefficient neither for dislocated atoms, nor for vacancies. The left-hand part of Eq. (38) presents the rate of the change of the volume of pores in a sample.

The number of dislocated atoms and vacancies created by the penetrating radiation are equal as a rule, that is $N_{ti} - N_{tv} = 0$. In this case
Eq. (38) is equivalent to

$$D_v k_{Tv} - D_i k_{Ti} \geq 0. \tag{39}$$

When the concentration of dislocated atoms and vacancies is much larger than the equilibrium concentration of vacancies, $k_{Tv} = -N_v u_{mv}/kT$ and $k_{Ti} = N_i u_{mi}/kT$. In this case one can see that the inequality, described by Eq. (39) is never possible. The left-hand part of Eq. (39) is always negative. This means that to prevent swelling we have to achieve circumstances when the concentration of dislocated atoms and the concentration of vacancies are close to the equilibrium concentration of vacancies. In this limit we have $k_{Tv} = N_{ev}(u_v - u_{mv})/kT$ and $k_{Ti} = N_i u_{mi}/kT$. Then the inequality, described by Eq. (39) is possible. To achieve this we have to introduce into a sample as many different sinks as possible (but not pores). This will help reducing the concentration of dislocated atoms and vacancies. Then the swelling of a material could be prevented provided $u_v$ is larger than $u_{mv}$ [4].

## 5. DISCUSSION

Experimental data on ohmic sintering [15, 18] show that during prolonged sintering, especially at higher temperatures, the peculiarities, caused by the driving forces arising due to bulk heating become more and more pronounced. This corresponds to the increasing contribution of stationary stage processes to the final structure and properties of sintered samples, and according to the theory, the contribution of the regarded driving forces are the largest at the isothermal regimes. After bulk heating sintering, according to the theory, samples have a specific defect structure: greater amount of pores of smaller sizes, which improves significantly mechanical properties of sintered materials.

The best way to prevent swelling of materials is to introduce as many sinks of vacancies as possible (grain boundaries, dislocations, etc., but not pores) and to use materials, where the enthalpy of vacancy creation is larger than the enthalpy of vacancy migration.

A mesoscopic approach, described in present paper, which had been used for years to solve sintering problems, also could help solving problems of nucleation, decomposition, and other problems in the theory of alloys. Described relaxation time technique could help simplifying calculations and give more clear physical picture.

Bulks heating driving forces, important for small size objects of microelectronics or/and heat evolving elements in nuclear reactors, are important also for the problems of nucleation and decomposition in these objects.

## 6. CONCLUSION

A general phenomenological theory of initial stages of ohmic sintering and final stages of a sintering process with and without bulk heating mechanisms is presented in this paper. The role of various sinks and sources of vacancies is analyzed. Mobile and immobile dislocations and their role in the sintering process are considered. The diffusional relaxation parameters for ensembles of grain boundaries, pores and dislocations are calculated. The role of the quality of the surface is discussed. Cycling and pulsing regimes of sintering are described. Specific driving forces of the bulk heating origin are analyzed in detail. The same approach is applied to the problem of swelling under penetrating radiation.